\documentclass[10pt,letterpaper]{article}
\usepackage{opex3,siunitx,amsmath}
\DeclareMathOperator{\sinc}{sinc}
\usepackage{amssymb}

\usepackage{color}
\usepackage{soul}

\setstcolor{red}

\begin{document}

\title{Near-field to far-field characterization of speckle patterns generated by disordered nanomaterials}

\author{Valentina Parigi$^{1}$, Elodie Perros$^1$, Guillaume Binard $^{2,3}$, C\'eline Bourdillon $^{2,3}$, Agn{\`e}s Ma{\^i}tre $^{2,3}$, R\'emi Carminati$^1$, Valentina Krachmalnicoff$^{1,*}$ and Yannick De Wilde$^{1,*}$}

\address{$^1$ESPCI ParisTech, PSL Research University, CNRS, Institut Langevin, 1 rue Jussieu, F-75005, Paris, France\\
$^2$Sorbonne Universit\'es, UPMC Univ Paris 06, UMR 7588, INSP, F-75005, Paris, France\\
$^3$ CNRS, UMR 7588, INSP, F-75005, Paris, France
}

\email{$^*$valentina.krachmalnicoff@espci.fr} \email{$^*$yannick.dewilde@espci.fr} 



\begin{abstract}

We study the intensity spatial correlation function of optical speckle patterns above a disordered dielectric medium in the multiple scattering regime. 
The intensity distributions are recorded  by scanning near-field optical microscopy (SNOM) with sub-wavelength spatial resolution at variable distances from the surface
 in a range which spans continuously from the near-field (distance  $  \ll \lambda $) to the far-field regime (distance $\gg \lambda $).
The non-universal behavior at sub-wavelength distances reveals the connection between the near-field speckle pattern and the internal structure of the medium.  
\end{abstract}


\section{Introduction}
Scanning near-field optical microscopy (SNOM) allows to recover the high spatial frequency components of an electromagnetic  field which are normally confined as evanescent waves at  sub-wavelength (sub-$\lambda$) distances from a surface in the so-called near-field region \cite{Novotny06}. The spatial resolution of the technique is determined by the nano-scale tip or aperture which is used to interact with the near field at nanometer separation from the sample. Since its first realization \cite{Pohl84}, SNOM has been mainly used to study samples, which consist of sub-wavelength structures placed at the surface of homogeneous substrates.   

The study of field correlations in speckle patterns produced by three-dimensional disordered materials has recently evolved towards the quest for non-universal properties, which could be connected with the morphology of the medium at sub-$\lambda$ scale \cite{Dogariu15}. It has been recently shown that the degree of mutual polarization for different points in speckle patterns depends on the scattering sample \cite{Broky10}, while spatial field correlations can reveal the relevant length scales of the medium morphology \cite{Carminati10} or the distance between point sources embedded in the sample \cite{Carminati15}.  In particular the speckle pattern in the near field of a scattering medium shows sub-$\lambda$ grain size which can be correlated with the sub-$\lambda$ structure of the medium. 

The connection between the field spatial correlation function in near-field speckle patterns and the structure of the medium has been analyzed in Ref.~\cite{Carminati10} using a simple model.
In this theoretical approach, the disordered medium is described by a fluctuating dielectric function $\epsilon(\textbf{r})=1+ \delta \epsilon(\textbf{r})$ considered as a random variable. The dielectric
function satisfies $\langle\delta\epsilon(\textbf{r})\rangle=0$ and $\langle \delta \epsilon(\textbf{r}) \delta \epsilon(\textbf{r}')\rangle=C(|\textbf{r}-\textbf{r}'|)$, with $C(R)$ a correlation function
with a width $\ell_{\epsilon}$ that characterizes the internal structure of the medium (here the brackets denote an ensemble average over the configurations of the random medium). In the regime
$\ell_{\epsilon} \ll \lambda \ll \ell$, with $\ell$ the scattering mean free path, the field correlation function above the exit surface of a three-dimensional scattering medium illuminated by a plane wave
can be calculated analytically. More precisely, the model predicts the behavior of the degree of spatial coherence 
$\gamma_E=\sum_k \langle E_k(\textbf{r})E_k(\textbf{r}')\rangle / \langle \vert E(\textbf{r}) \vert ^2 \rangle$, taken as a measure of the overall spatial correlation of the vector field, the subscript k denoting a vectorial component. Its width defines the
spatial coherence length. In the experiments presented in the present study, we have measured the normalized intensity correlation function $\gamma_I=\langle I(\textbf{r})I(\textbf{r}')\rangle / \langle \vert I(\textbf{r}) \vert ^2 \rangle$, whose
width is the average size of a speckle spot. In a fully developed speckle pattern, the field exhibits a Gaussian statistics, so that $\gamma_I = 1 + |\gamma_E|^2$ \cite{Dogariu15}, and we will identify the size of a speckle spot with the spatial coherence length. The model predicts three different regimes for the speckle patterns measured in a plane at a distance $z$ above the exit surface of the medium. For $z\gg\lambda$ (far field), the behavior of fully developed speckle pattern is universal, and the speckle spot size depends only on the wavelength and the illumination geometry (for plane-wave illumination the size is $\lambda/2$). On the contrary in the near field zone $z<\lambda$, the speckle loses its universal behavior, and the speckle spot size decreases linearly with $z$. Finally, in the extreme near-field regime for which $z\sim\ell_{\epsilon}\ll \lambda$, the speckle spot size reaches an asymptotic value on the order of $\ell_{\epsilon}$, that reflects the internal structure of the disordered medium.
As a consequence, in this extreme near-field regime, the structural properties of the medium are encoded in a two-dimensional image of the speckle pattern, provided that the measurement is able to resolve the speckle grains with sufficient spatial resolution.
 
The non-universal behavior of intensity speckle pattern in the near field has already been put forward in some previous experimental studies \cite{Apostol03,Emiliani03,Apostol04,Apostol05}, which in particular have shown speckle spot sizes below the diffraction limit at sub-wavelength distance from the surface.  In this paper, we explore for the first time the continuous transition from the far field to the extreme near field using SNOM, leading to a measurement of the speckle spot size for distances to the medium surface ranging from a few nanometers to several tens of micrometers. The behavior of the speckle spot size versus the distance reveals different regimes, in excellent agreement with the predictions of the model described above.
Besides the possibility of disclosing the relevant length scales of nanostructured samples, the investigation of fundamental lower limits in speckle grain size is an important issue for imaging techniques which rely on spatial field or intensity correlations, focusing or wavefront control in disordered media \cite{Weaver01,Derode02,Lerosey07,Vellekoop07,Popoff10}.  
As the spatial resolution is driven by the speckle grain size, the non-universal behavior in the near-field range could push the resolution beyond the diffraction limit.

\section{Experiment}

\begin{figure}[htbp]
\centering\includegraphics[width=10cm]{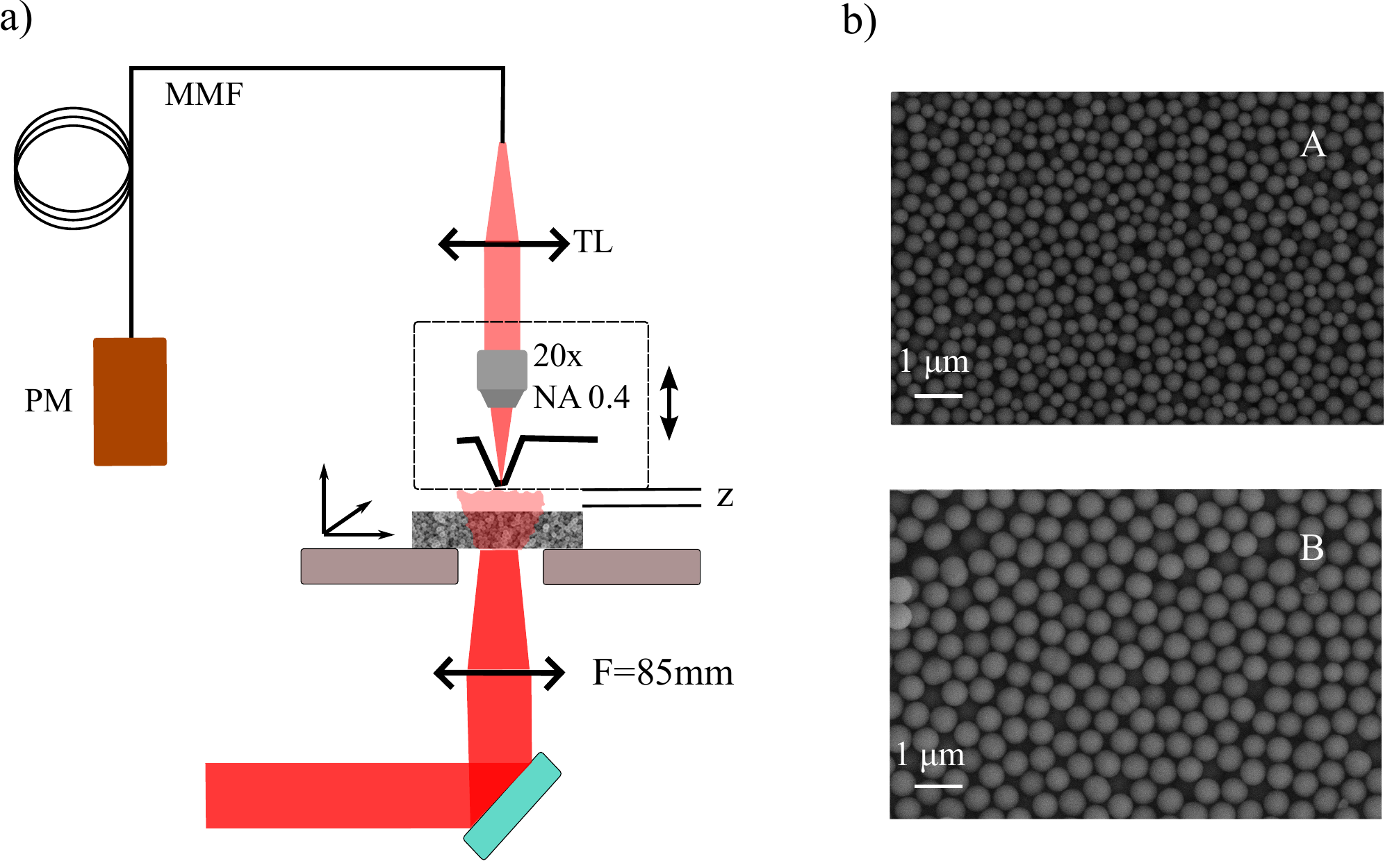}
\caption{a) Scheme of the setup. The laser beam is focused on the sample down to a diameter of  $\sim$\SI{40}{\micro\metre}. The sample is placed on a three-axis stage which allows scanning in the x-y plane during the acquisition. The z-stage is used in contact mode to follow the surface topography with the pyramid when the electronic feedback is in operation.   The  light above the sample surface passing through the $\sim$ \SI{150}{\nano\metre} wide hole in the pyramid is collected by the confocal microscope (objective $20$X, NA $0.4$). It is injected by means of the microscope tube lens (TL, focal distance 200 mm) into a multimode fiber (MMF) connected to a photomultiplier (PM) which measures the light  intensity. The $z$ distance between the pyramid and the sample surface is  controlled by the sample-stage unit up to \SI{8}{\micro\metre} and then by jointly moving the cantilever and the objective.  b) Scanning electron microscope images of the surfaces of the two analyzed samples.}
\label{setup}
\end{figure}

Our experimental apparatus, schematized in Fig. \ref{setup}a, makes use of a commercial SNOM (WITec GmbH alpha300S). It is conceived as an atomic force microscope (AFM) combined with an optical confocal microscope with an objective  of magnification 20X and numerical aperture $NA=0.4$. The AFM cantilever includes a hollow metal coated pyramidal tip  with a \SI{150}{\nano\metre} wide aperture carved at the apex of the pyramid. It is set under the objective, the apex being located at focal distance. 
The light passing through the aperture and collected by the objective is convoyed to a multimode fiber which is connected with a photomultiplier. The sample is placed under the tip on a  multi-axis piezoelectric translation stage and is illuminated from below by a monochromatic field at $\lambda=$ \SI{633}{\nano\metre}, produced by a He-Ne laser with a maximum power of 20 \SI{}{\milli\watt}. The setup, which normally has a second objective in order to focus the beam in a limited region under the sample, is slightly modified by replacing the objective with a plano-convex lens with $F=$\SI{85}{\milli\metre}.  This produces a beam that illuminates the sample on a region of diameter $D=40\,\mu$m. 

The apparatus can work in two different configurations. In the first one (contact mode), the pyramid is kept in contact with the surface of the sample by an electronic feedback, while the transducer  moves the sample in the x-y plane. In this configuration, we simultaneously acquire  the topographic (AFM) image of the surface and the spatial intensity distribution of the electromagnetic field above the sample, measured as the number of counts due to photons arriving on a photomultiplier (PM).   The resolution is driven by the diameter of the nanoaperture. In the second configuration (constant height mode), the  separation between the sample and the apex of the hollow pyramid, $z$, is fixed and the x-y scan is performed without turning on the feedback loop, while acquiring only the intensity distribution. The $z$ distance is adjusted by means of the piezo-transducer itself which can retract the sample up to $z=$ \SI{8}{\micro\metre} below the apex of the hollow pyramid. Larger distances can be reached by rising the microscope above the sample, while simultaneously moving the pyramid and the objective. 

We investigated two different samples consisting of silica spheres in a disordered arrangement. Two images of their surfaces, recorded with a scanning electron microscope (SEM)  are shown in Fig. \ref{setup}b. The two samples are composed of several layers of beads with different average diameters: statistics performed on SEM images give an average diameter of $\phi = \SI{276}{\nano\metre}$  and $\phi = \SI{430}{\nano\metre}$ for sample A and B respectively. 
Multiple scattering of the incident laser beam transmitted through the sample generates speckle patterns without residual ballistic component. 

We typically perform scans of \SI{10}{\micro\metre}$\times$\SI{10}{\micro\metre} consisting of $256\times 256$ points. In case of large distances from the sample,  as we found larger speckles grain sizes,  we took \SI{20}{\micro\metre}$\times$\SI{20}{\micro\metre} scans with $256\times 256$ points in order to maintain a significant number of grains within the scanned area. 

\begin{figure}[htbp]
\centering\includegraphics[width=9cm]{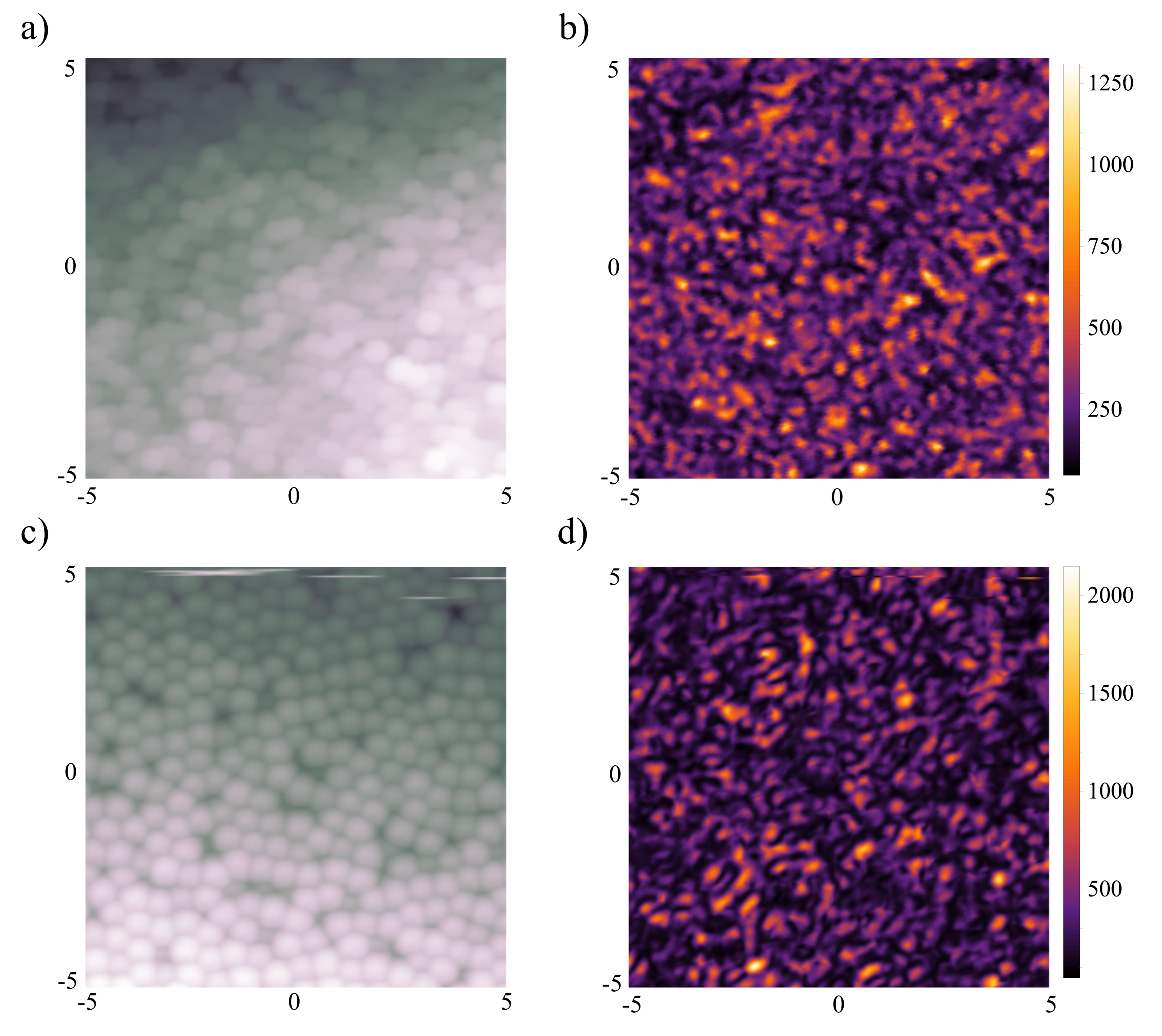}
\caption{Two scans in contact mode for the two samples. a) Topography,  and b) intensity images measured in one zone of sample A. c) and d) the same for sample B.  The vertical and horizontal scales units are \SI{}{\micro\metre}, while the values on the colorbar are number of counts of the PM.  }
\label{contact}
\end{figure}

\begin{figure}[htbp]
\centering\includegraphics[width=10cm]{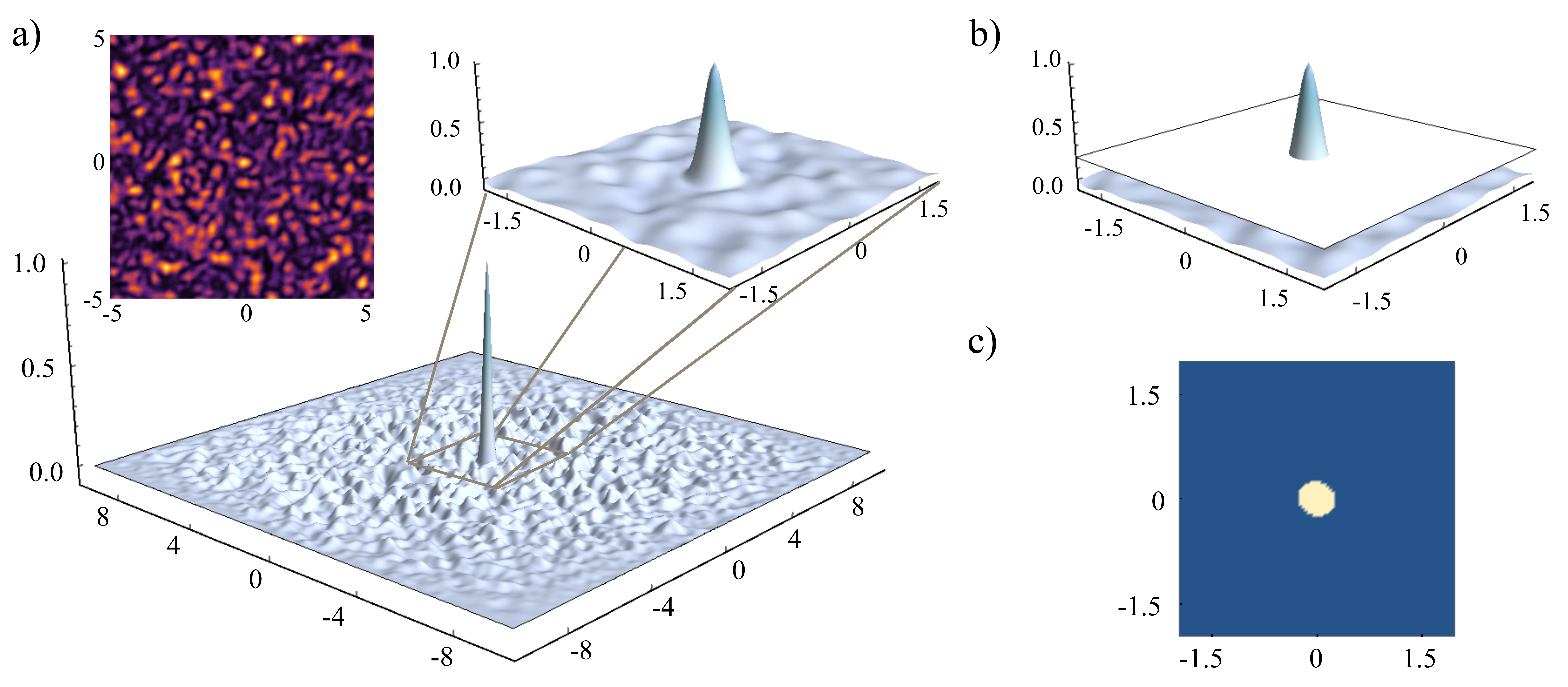}
\caption{Evaluation of speckle grain size. a)
On upper part left: typical intensity speckles pattern. Bottom: its autocorrelation  with a zoom of the central region (upper right). b) The autocorrelation is cut with a plane at a height of 0.2. c) The surface resulting from the cut: the speckle size $\delta$ is obtained as the average radius of the surface. Scales units on the plane are \SI{}{\micro\metre}.}
\label{cut}
\end{figure}

Figure \ref{contact} shows two typical results obtained in contact mode on both samples. The left column of the figure shows the topographic images, while the right column shows the  intensity maps, which have been acquired at the same time as the topography. They correspond to fully developed speckle patterns generated by the He-Ne laser transmitted through the disordered volume of the sample. As can be noticed in Fig. \ref{contact} the intensity patterns are not direct images of the beads distribution at the surface, since they result from the multiple scattering process into the volume. Looking at the scan scale one can already notice that the typical bright grain size is below the micrometer range. 
 In order to get quantitative measurements, we define  the speckle grain size as the half-width of the base of the autocorrelation peak. With such definition the half-width of the autocorrelation function in the far-field, in the case of a fully developed speckle generated under plane wave illumination, is  $\delta_0=\lambda/2$, corresponding to the first zero value of $\gamma_I(\textbf{r},\textbf{r}')$. In fact, the autocorrelation has the form of $\gamma_I(\textbf{r},\textbf{r}')= \sinc (k_0 \vert  \textbf{r}-\textbf{r}' \vert)^2= \sinc (k_0  \rho )^2$, with $k_0=2\pi/\lambda$.
 Data analysis follows the procedure depicted in Fig. \ref{cut}: we calculate the normalized autocorrelation function of the intensity scan  and we cut it with a plane at $0.2$.  Note that the threshold value of $0.2$ has been chosen rather than $0$ in order to get rid of noise. Then we keep as grain size the average radius $\delta$ of the resulting surface $S$ by calculating $\delta =\sqrt{S/\pi}$. 
 
\begin{figure}[htbp]
\centering\includegraphics[width=12cm]{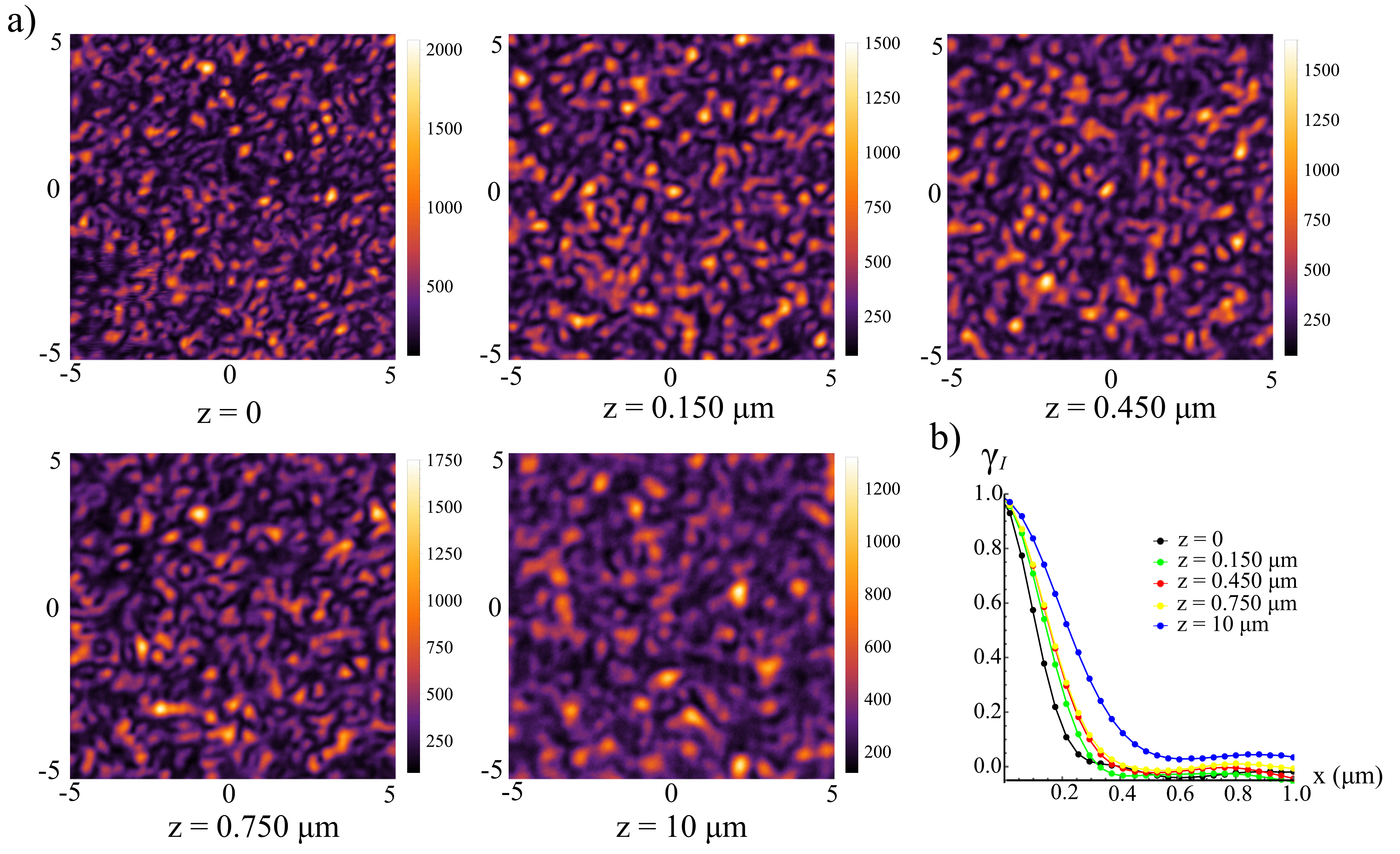}
\caption{a) Series of intensity speckle patterns in one zone of the sample B at different distances from the surface ranging from the near-field to the far-field regime. The vertical and horizontal scales units are \SI{}{\micro\metre}, while the values on the colorbar are number of counts of the PM.  b) Sections along the x direction of the autocorrelation function obtained from the five patterns.} 
\label{speckles}  
\end{figure}

\section{Results}
\begin{figure}[htbp]
\centering\includegraphics[width=10cm]{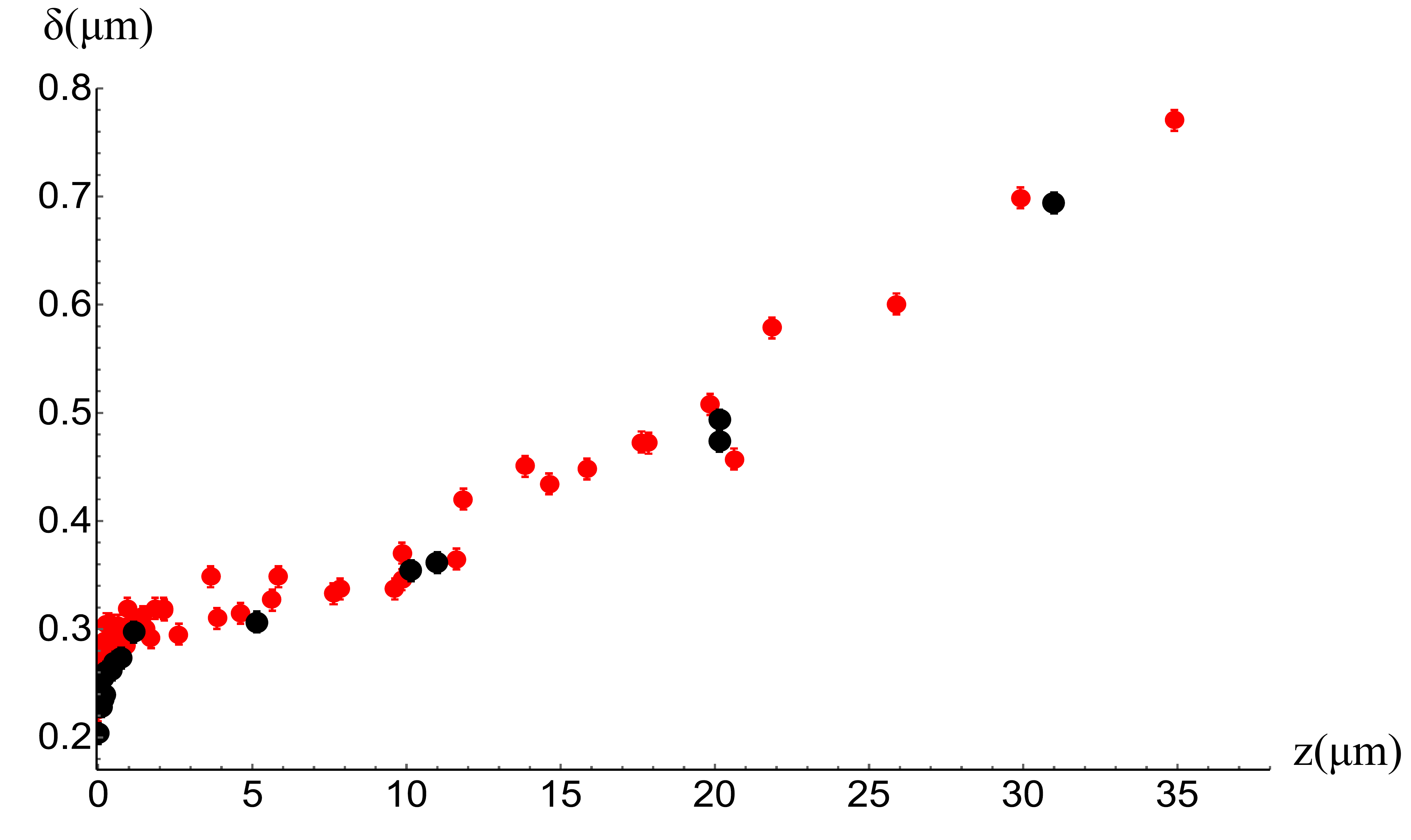}
\caption{Measurement of autocorrelation size from the near-field to the far-field regime in the case of sample A (red points) and sample B (black) points. The curves are composed by several series of measurements varying $z$ in different zones of the samples.}
\label{fullcurve}
\end{figure}

 Figure \ref{speckles}a shows a typical series of measurement, obtained by changing the distance z from a given region of the sample B. The first two measurements, for $z=$\SI{0}{\micro\metre} and $z=$\SI{0.150}{\micro\metre}, belong to the near field regime $z<\lambda/2$. The two following measurements, at distances  z close to  $\lambda=$\SI{0.633}{\micro\metre}, show a similar speckle size, while the last picture at larger distance shows a clearly increased grains size.  Figure \ref{speckles}b shows the section of the autocorrelation function for each distance and allows us to measure the speckle grain size, obtaining the following series $\delta= (0.208,0.255,0.262,0.273,0.354)$ \SI{}{\micro\metre}.
In order to confirm this behavior and to study the transition between the different regimes, we performed several series of measurements in different regions of both samples, obtaining the curve in Fig. \ref{fullcurve}, which spans between $z=$0 and $z=$\SI{30}{\micro\metre}. This is the main result of this work.
It constitutes, to the best of our knowledge, the first measurement of spatial correlations in speckle patterns continuously spanning from the near field to the far field, maintaining the same experimental configuration.   

In the far field, $z\gg\lambda$, the spatial properties of fully developed speckle patterns are universal \cite{Goodmanbook,Dogariu15}.  Being $D$ the transverse size of the beam illuminating the sample, for distances $z \ll D$ the beam can be approximated as an infinite plane wave. Therefore, the autocorrelation function has the typical sinc form, previously  mentioned, corresponding to a grain size which only depends  on the illuminating wavelength ($\delta=\lambda/2$).  For distances to the sample of the same order as or larger than the beam size, the autocorrelation function has a more complex, although well known, behavior \cite{Goodmanbook} and the speckle grain size varies as $\delta = \lambda z/ D$.  
 We observed both regimes. Since $D\simeq 40\,\mu$m, for $\lambda/2 \leq z \leq$ \SI{10}{\micro\metre} the grain size is mainly constant and not so far from the $\delta=\lambda/2$ value.  For distances $z>$ \SI{10}{\micro\metre} the grain size starts to depend on the beam geometry and the distance following the $\delta = \lambda z/ D$ law.
Fitting on data at distances $z>$ \SI{10}{\micro\metre} we found $D=$\SI{38.6}{\micro\metre} and $D=$\SI{39.3}{\micro\metre} for sample A and B respectively, which are compatible with the estimated beam diameter.
 
We further studied the far field range in order to check the reliability of our experimental apparatus. At distances $z>\lambda$, where the super-resolution given by the nanoscale probe is not strictly necessary, we collected the speckle patterns both in SNOM configuration (by collecting light through the aperture of the cantilever) and in far-field microscope configuration. The second one is obtained by removing the cantilever and using the setup as a classical confocal microscope in which the fiber core constitutes the pinhole. In that case, in order to achieve a spatial resolution of $\approx\lambda/2$, we replaced the 20X objective by a 100X, $NA=0.9$ objective and we used a single mode fiber (\SI{5}{\micro\metre} core) instead of the multimode fiber. In this large distance regime, we found no significant difference between images obtained in the confocal microscope and in the SNOM configuration.

\begin{figure}[htbp]
\centering\includegraphics[width=12cm]{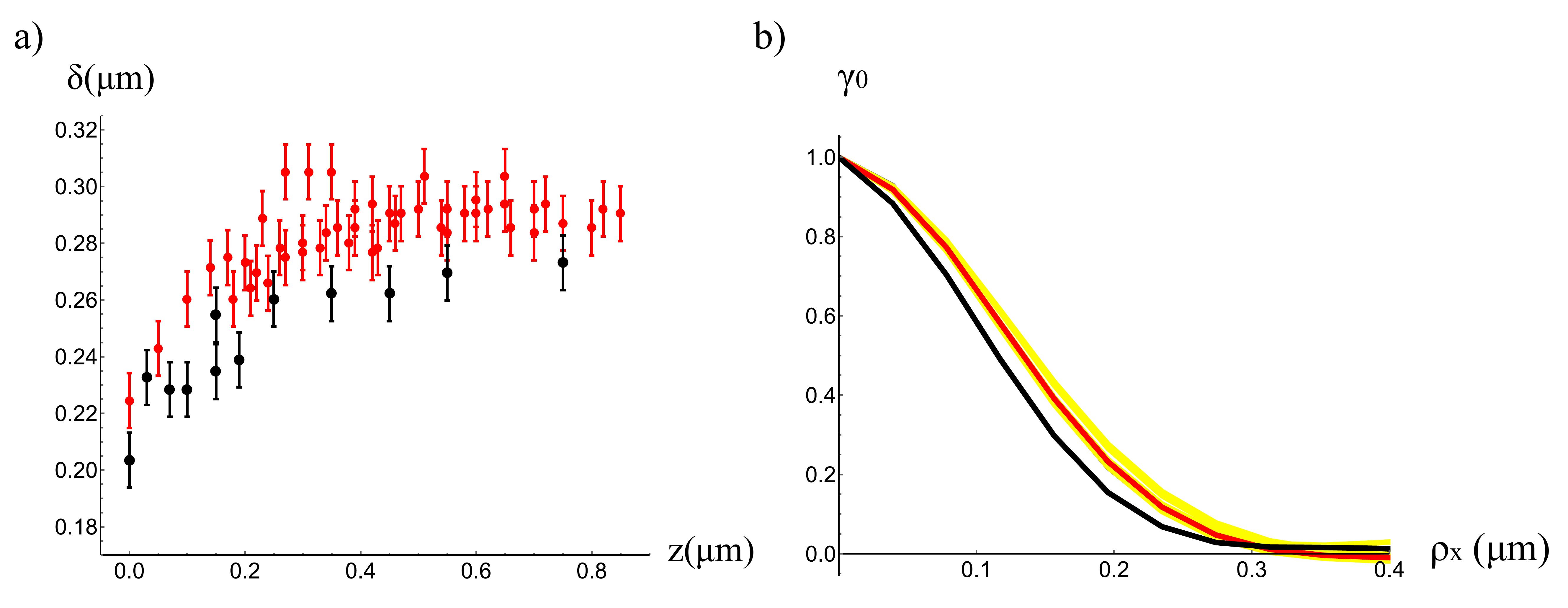}
\caption{a) Near-field subset of data in Fig.\ref{fullcurve}. The two points at $z=0$ result from averaging over 9 measurements in the case of sample A (red) and over 3 measurements in case of sample B (black). b) Section along the x direction of the autocorrelation functions for  $z=0$. Red curve: average over the measurements of sample A,  black curve: average over the measurements of sample B, yellow curves: the 9 measurements for sample A at $z=0$.  }
\label{curveNF}
\end{figure}

The $z<\lambda$ range is the most interesting one: it contains the far-field to near-field transition, which shows up at $z \sim \lambda/2$, as it can be observed in Fig. \ref{curveNF} a, which is a zoom of the near-field zone of the curve in Fig. \ref{fullcurve}.
Below  $z = \lambda/2$ the speckle size loses its universal nature and starts decreasing by approaching the surface.  Data acquired for samples A and B are close to each other and have a linear dependence with $z$, as predicted by theory \cite{Carminati10}. In the extreme near field, i.e. in the limit of $z\to 0$, the two  data sets are clearly separated in the limit of the experimental errors,  and we measure $\delta = 225 \pm 10$~nm and $\delta = 204 \pm 10$~nm for sample A and B respectively. 

The $z=0$ case corresponds to the configuration where the hollow pyramid is kept in contact by means of the electronic feedback. This is hence the most repeatable configuration and we took several measurements: the first point of the two curves is the average value over 9 and over 3 measurements for the sample A and B respectively. The error bar in the plot is calculated as the maximal deviation over the 9 measurements for the sample A. All the other points in the curves are calculated from one single acquisition of speckle pattern at the given distance.
In order to evaluate the error in the estimation of the $z$ value we have to take into account that, once left the contact configuration with the feedback locking, the cantilever undergoes possible  drifts due to any air flow or thermal and mechanical disturbance. 
Even if vibrations and air flow are reduced  by opportune isolation, thermal drifts are unavoidable due to the proximity of electronic parts which dissipate heat within the enclosure of the SNOM. Experimental tests show a thermal drift of the order of \SI{200}{\nano\metre} per hour, which means \SI{20}{\nano\metre} in a $6$-minutes long acquisition. For this reason acquisitions are taken one after the other and by always changing $z$ in the same direction (either increasing or decreasing the value). Thermal drift is anyway difficult to control and we are not exempt to possible fluctuations, which is the reason why we did not add horizontal error bars in Fig. \ref{fullcurve}.  

We can reasonably assume that the two average grain sizes at $z=0$ have been estimated  at a distance from the surface which is smaller than the typical length scale $\ell_{\epsilon}$ describing the fluctuation of the dielectric function of the medium in a simple theoretical approach \cite{Carminati10}. This length should be connected with the nano-scale structure of the sample, which in our case should be driven by the size of silica beads and their arrangement within the sample. The presence and the size of air gaps between the beads depend on parameters such as the homogeneity of the bead sizes,  homogeneity of the beads distribution inside the medium, etc. The average size of the beads and their dispersion around the central value could be inferred from the SEM images of the surface, and it is also possible to have an idea on the arrangement of beads on the surface. However, it is not possible to know how the different layers of beads are arranged in the volume, so it is not possible to predict  the correlation length  for the two samples. Nevertheless, from the different values of $\delta$ found experimentally on the two samples at $z=0$, i.e. in the extreme near-field regime, one can clearly affirm that the two samples have a different micro-structure, independently of the  different average bead diameters,  proving that this method reveals material properties that would remain hidden with other optical microscopy techniques.

\section{Conclusion}
In conclusion we have measured the intensity spatial correlation function in speckle patterns produced by disordered dielectric materials, made of sub-micrometer beads. 
We have performed series of measurements spanning from the near-field ($z\ll\lambda$) to the far-field ($z\gg\lambda$) regime using a scanning near-field optical microscope. The averaged speckle spot size,
defined as the width of the intensity correlation function, has been measured from the extreme near field to the far field on an unprecedented distance range, revealing different regimes
in excellent agreement with theoretical predictions. In the near field regime, the non-universal behavior of speckle patterns has been clearly demonstrated, showing the ability of near-field speckles to
discriminate between materials with different internal microscopic structures.
  
\section*{Acknowledgment}
 The authors thank C. Aydin and L. Greusard for help in setting up the experiment. This work was supported by LABEX WIFI (Laboratory of Excellence ANR-10-LABX-24) within the French Program ``Investments for the Future" under reference ANR-10- IDEX-0001-02 PSL* and by the French National Research Agency (ANR ``CALIN" and ANR ``GOSPEL").


\begin{thebibliography}{99}
\bibitem{Novotny06} L. Novotny and B. Hecht, ``Principles of Nano-Optics,'' 2nd ed. (Cambridge University Press, New York, 2006).
\bibitem{Pohl84} D. Pohl, W. Denk, and M. Lanz, ``Optical stethoscopy: Image recording with resolution $\lambda/20$,'' Appl. Phys. Lett.{\bf 44}, 651 (1984).
\bibitem{Dogariu15} A. Dogariu and R. Carminati, ``Electromagnetic field correlations in three-dimensional speckles,'' Physics Report {\bf 559}, 1-29 (2015).
\bibitem{Broky10} J. Broky and A. Dogariu, ``Complex degree of mutual polarization in randomly scattered fields,'' Optics Express  Vol. 18, Issue 19, pp. 20105-20113 (2010).
\bibitem{Carminati10} R. Carminati, ``Subwavelength spatial correlations in near-field speckle patterns,'' Phys. Rev. A {\bf 81}, 053804 (2010).
\bibitem{Carminati15}R. Carminati, G. Cwilich, L. S. Froufe-P{\'e}rez, and J. J. S{\'a}enz, ``Speckle fluctuations resolve the interdistance between incoherent point sources in complex media,''Phys. Rev. A 91, 023807 (2015).
\bibitem{Apostol03} A. Apostol and A. Dogariu , ``Spatial Correlations in the Near Field of Random Media,'' Phys. Rev. Lett. {\bf 91}, 093901 (2003).
\bibitem{Apostol04} A. Apostol and A. Dogariu , ``First- and second-order statistics of optical near fields,'' Optics Lett. {\bf 29}, 235-237 (2004).
\bibitem{Emiliani03} V. Emiliani, F. Intonti, M. Cazayous, D. S.Wiersma, M. Colocci, F. Aliev and A. Lagendijk, , ``Near-Field Short Range Correlation in Optical Waves Transmitted through Random Media,'' Phys.
Rev. Lett. {\bf 90}, 250801 (2003).
\bibitem{Apostol05} A. Apostol and A. Dogariu , ``Non-Gaussian statistics of optical near-fields,'' Phys. Rev. E {\bf 72}, 025602 (2005).
\bibitem{Weaver01} R. L. Weaver and O. I. Lobkis, ``Ultrasonics without a Source: Thermal Fluctuation Correlations at MHz Frequencies,'' Phys. Rev. Lett. \textbf{87}, 134301
(2001).
\bibitem{Derode02} A. Derode, A. Tourin, and M. Fink, ``Time reversal versus phase conjugation in a multiple scattering environment,'' Ultrasonics \textbf{40}, 275 (2002).
\bibitem{Lerosey07} G. Lerosey, J. de Rosny, A. Tourin, and M. Fink, ``Focusing beyond the diffraction limit with far-field time reversal''  Science \textbf{315}, 1120 (2007).
\bibitem{Vellekoop07} I. M. Vellekoop and A. P. Mosk, ``Focusing coherent light through opaque strongly scattering media,'' Opt. Lett. \textbf{32}, 2309 (2007). 
\bibitem{Popoff10}S. M. Popoff, G. Lerosey, R.  Carminati, M. Fink, A. C. Boccara, and S. Gigan, ``Measuring the Transmission Matrix in Optics: An Approach to the Study and Control of Light Propagation in Disordered Media,'' Phys. Rev. Lett. \textbf{104}, 100601 (2010).
\bibitem{Goodmanbook} J. Goodman, \textit{Speckle Phenomena in Optics}  (Roberts \&  Company Publishers 2010)

\end{thebibliography}
\end{document}